\RequirePackage{etoolbox}
\csdef{input@path}%
{%
 {sty/},
 {img/},
}
\documentclass{elsevierbook}

\usepackage{biblatex}
\usepackage{soul}
\usepackage{dirtree}
\usepackage{multirow}
\usepackage{xcolor}
\usepackage{algorithm}
\usepackage{algpseudocode}
\usepackage{booktabs}
\usepackage{multicol}
\usepackage{pifont}
\usepackage{float}
\usepackage{dirtytalk}
\DeclareUnicodeCharacter{0301}{\'{e}}

\begin{document}

\Frontmatter

\Mainmatter
  \begin{frontmatter}

\chapter{CapsNet for Medical Image Segmentation}\label{chap1}

\begin{aug}
\author[addressrefs={ad1}]%
  {\fnm{Minh}   \snm{Tran}}%
\author[addressrefs={ad1}]%
  {\fnm{Viet-Khoa}   \snm{Vo-Ho}}%
\author[addressrefs={ad2}]%
  {\fnm{Kyle}   \snm{Quinn}}%
\author[addressrefs={ad3}]%
  {\fnm{Hien}   \snm{Nguyen}}%
\author[addressrefs={ad1}]%
  {\fnm{Khoa}   \snm{Luu}}%
  \author[addressrefs={ad1}]%
  {\fnm{Ngan}   \snm{Le}}%

\address[id=ad1]%
  {Department of Computer Science \& Computer Engineering, University of Arkansas, Fayetteville, USA 72703 }%
\address[id=ad2]%
  {Department of Biomedical Engineering, University of Arkansas, Fayetteville, USA 72703 }%
 \address[id=ad3]%
  {Department of Electrical \& Computer Engineering, Houston, TX 77204}%
\end{aug}

\begin{abstract}
Convolutional Neural Networks (CNNs) have been successful in solving tasks in computer vision including medical image segmentation due to their ability to automatically extract features from unstructured data. However, CNNs are sensitive to rotation and affine transformation and their success relies on huge-scale labeled datasets capturing various input variations. This network paradigm has posed challenges at scale because acquiring annotated data for medical segmentation is expensive, and strict privacy regulations. Furthermore, visual representation learning with CNNs has its own flaws, e.g., it is arguable that the pooling layer in traditional CNNs tends to discard positional information and CNNs tend to fail on input images that differ in orientations and sizes. Capsule network (CapsNet) is a recent new architecture that has achieved better robustness in representation learning by replacing pooling layers with dynamic routing and convolutional strides, which has shown potential results on popular tasks such as classification, recognition, segmentation, and natural language processing. Different from CNNs, which result in scalar outputs, CapsNet returns vector outputs, which aim to preserve the part-whole relationships. In this work, we first introduce the limitations of CNNs and fundamentals of CapsNet. We then provide recent developments of CapsNet for the task of medical image segmentation. We finally discuss various effective network architectures to implement a CapsNet for both 2D images and 3D volumetric medical image segmentation. 
\end{abstract}
\begin{keywords}
\kwd{Capsule Network}
\kwd{CapsNet}
\kwd{Medical Image}
\kwd{Segmentation}
\end{keywords}

\end{frontmatter}

\section{Convolutional Neural Networks: Limitations}
Despite outperforming in various computer vision tasks, CNNs ignore the geometrical relationships of objects. As a result, CNNs are sensitive to image rotation and affine transformation, which are not present in the training data. Recent works have shown that small translations or rescalings of the input image can drastically change the network’s performance. To address such limitations in CNNs, their generalization relies on a large-scale training data, which captures various input variations such as rotations and viewpoint changes. In this section, we first report some quantified analyses of this phenomenon. We then analyze how the CNNs architectures do not produce invariance.   

\begin{figure}[h]
    \centering
    \includegraphics[width=\textwidth]{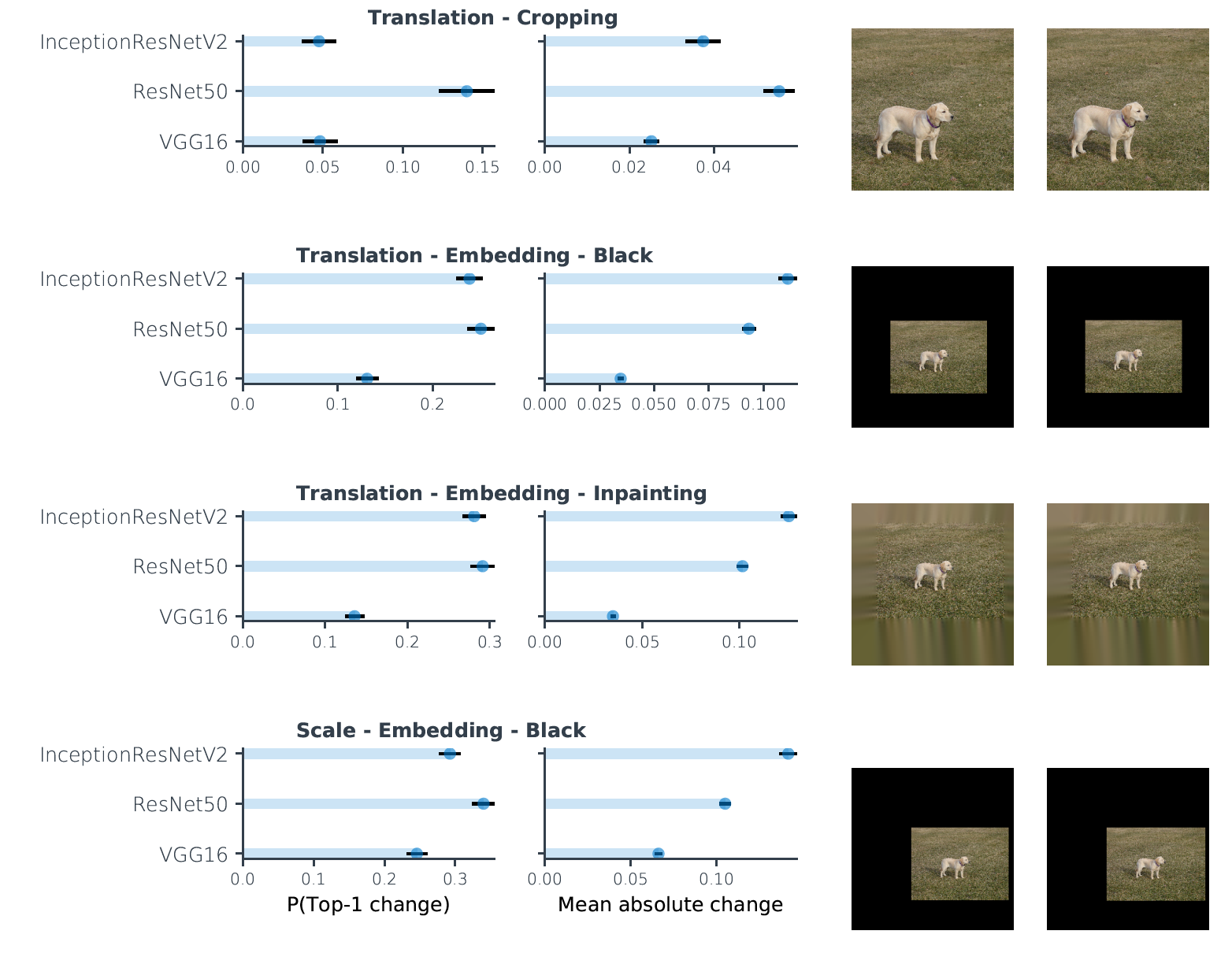}
    \caption{Quantitative analysis on three Keras networks with four different protocols on 1000 randomly chosen images from the ImageNet validation set. Image is from \cite{azulay2020deep}.}
    \label{fig:CNN_Compare}
\end{figure}

The lack of invariance of modern CNNs to small image deformation was reported in \cite{engstrom2018rotation, azulay2018deep, zhang2019making, azulay2020deep}. Take \cite{azulay2020deep} as an instance. Azulay and Weiss selected three different CNN architectures VGG16, ResNet50, InceptionRes- NetV2 from Keras package and three other different CNN architectures VGG16, ResNet50, DenseNet121 from Pytorch package. They tested on 1000 images with four different protocols including crop, Translation - Embedding - Black, Translation - Embedding - Inpainting, Scale - Embedding - Black to systematically quantify the effect of invariance to CNNs. In the first protocol, a random square is randomly chosen within the original image and resize the square to be 224x224. In the second protocol, the image is downsampled so that its minimal dimension is of size 100 while maintaining aspect ratio, and embed it in a random location within the 224x224 image, while filling in the rest of the image with black pixels. They then shift the embedding location by a single pixel, again creating two images that are identical up to a shift by a single pixel. In the third protocol, they repeat the embedding experiment but rather than filling in the rest of the image with black pixels we use a simple inpainting algorithm (each black pixel is replaced by a weighted average of the non black pixels in its neighborhood). The fourth protocol is similar to the second protocol, but rather than shifting the embedding location, they keep the embedding location fixed and change the size of the embedded image by a single pixel. They used \say{P(Top-1 change)} and \say{mean absolute change} (MAC) to measure the network sensitivity. The first metric \say{P(Top-1 change)} is invariant to any monotonic transformation of the output of the final layer of the network, while the second one tells us the possibility that changes in the top-1 prediction are due to very small differences between the most likely class and the second most likely class. The quantitative analysis on three Keras networks is shown in Fig.\ref{fig:CNN_Compare} which indicates that CNNs are not fully translation invariant.

Furthermore, in a CNN, a pooling layer works as a messenger between two layers and transfers the activation information from one layer to the next layer. By doing that, a pooling layer can indicate the presence of a part, but is unable to capture the spatial relation among the parts. Clearly, a pooling operation (e.g. max-pooling) does not make the model invariant to  viewpoint changes. Moreover, each filter of convolutional layers works like a feature detector in a small region of the input features. When going deeper into a network, the detected low-level features are aggregated and become high-level features that can be used to distinguish between different objects. The higher-level features in a CNN are built as a weighted sum of lower-level features. Thus, geometric relationships among features are not taken into account.

Over the years, different techniques have been developed to tackle the aforementioned limitations in CNNs. Most common solutions including data augmentation techniques, which increase the data size to include transformations.  

\section{Capsule Network: Fundamental}
Inspired by how a visual image was constructed from geometrical objects and matrices that represent relative positions and orientation, Hinton \cite{hinton2011transforming} proved that preserving hierarchical pose relationships among object parts is important to correctly classify and recognize an object. This is also known as inverse graphics, similar to how the brain recognizes the object. To address the limitations in CNNs, Hinton \cite{hinton2011transforming} proposed to replace neuron's scalar output which only represents the activation of replicated feature detectors with vector output (a.k.a. capsule). Each capsule will learn a visual entity (e.g. a part of an object). The output of each capsule is presented by a vector that contains both the probability that this entity is present and a set of "instantiation parameters" that captured different properties of the entity

CapsNet \cite{sabour2017dynamic} was proposed to address the above intrinsic limitations of CNNs. CapsNet strengthens feature learning by retaining more information at the aggregation layer for pose reasoning and learning the part-whole relationship. Different from CNNs that contain a backbone network to extract features, several fully connected layers, and N-way Soft-Max produces the classification logits, a CapsNet contains more complex five components as follows:

\begin{itemize}
    \item Non-shared transformation module: the primary capsules are transformed to execute votes by non-shared transformation matrices.
    \item Dynamic routing layer: to group input capsules to produce output capsules with high agreements in each output capsule.
    \item Squashing function: to squash the capsule vectors' lengths to the range of [0, 1).
    \item Marginal classification loss: to work together with the squashed capsule representation.
    \item Reconstruction loss: To recover the original image from the capsule representations.
\end{itemize}

\begin{figure}[h]
    \centering
    \includegraphics[width=\textwidth]{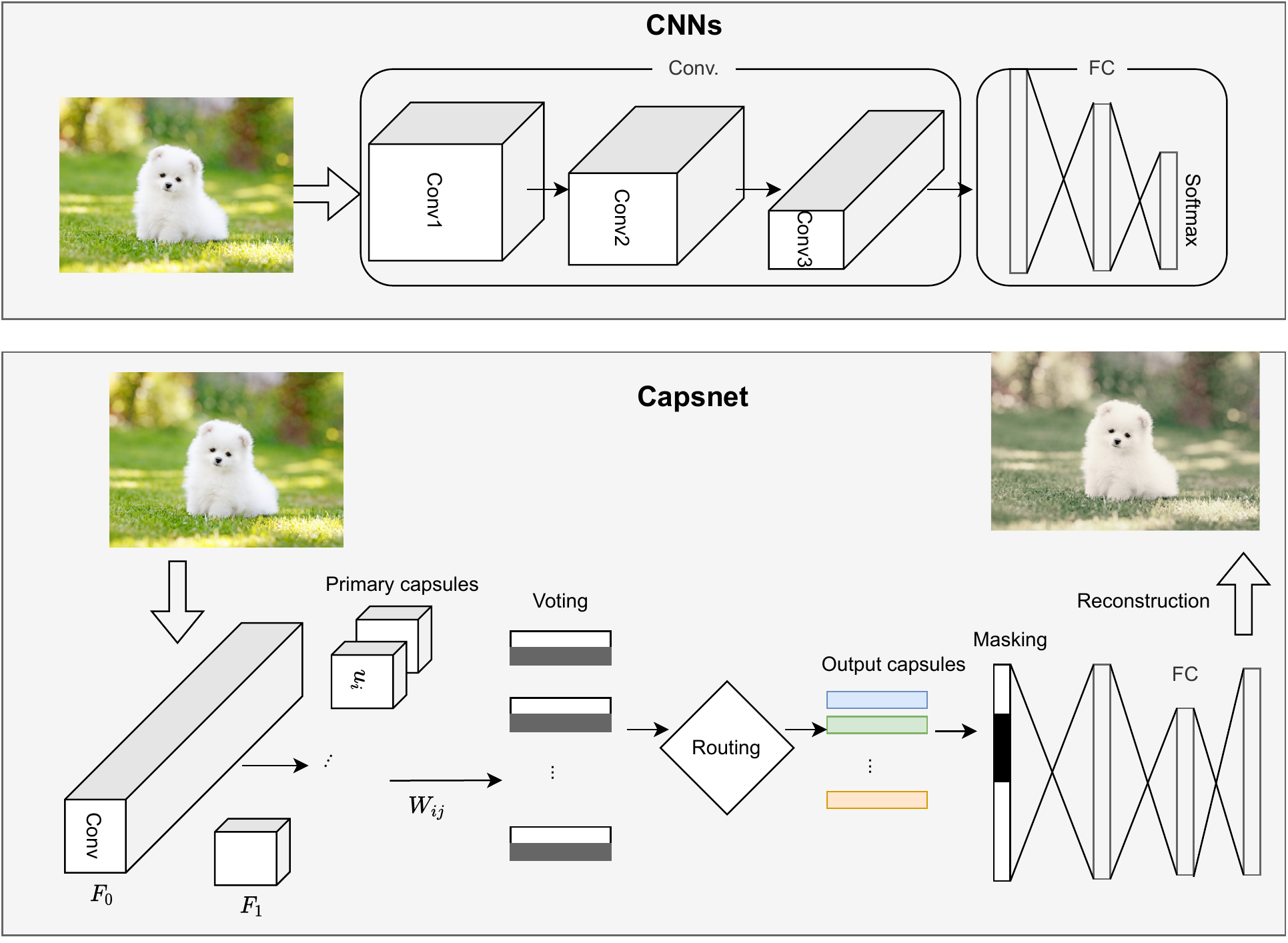}
    \caption{Network architecture comparison between CNNs and CapsNet.}
    \label{fig:CNN-Caps-Compare}
\end{figure}

\begin{figure}[h]
    \centering
    \includegraphics[width=\textwidth]{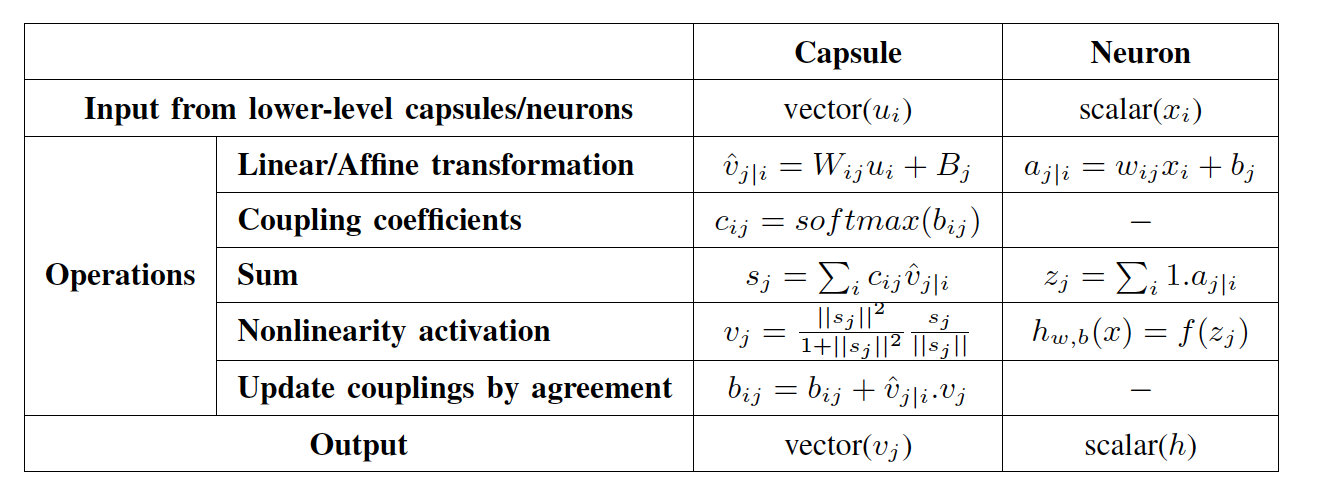}
    \caption{Operators comparison between CNNs and CapsNet.}
    \label{ig:CNN-Caps-Compare2}
\end{figure}

The network architecture comparison between CNNs and CapsNet is shown in Fig. \ref{fig:CNN-Caps-Compare} whereas the operators, input, and output comparison is given in Fig. \ref{ig:CNN-Caps-Compare2}. In CNNs, each filter takes in charge of feature detector in a small region of the input features and as we go deeper. The detected low-level features are aggregated and become high-level features that can be used to distinguish between different objects. However, by doing so, each feature map only contains information about the presence of the feature, and the network relies on fix learned weight matrix to link features between layers. It leads to the problem that the model cannot generalize well to unseen changes in the input \cite{alcorn2019strike}. In CapsNet, each layer aims to learn a part of an object together with its properties and represent them in a vector. The entity of previous layer represents simple objects whereas the next layers represents complex objects through the voting process.

Let denote $F_0\in\mathbb{R}^{C \times W\times H}$ the visual feature map which is extracted by a few convolutional layers. It then is reshaped as primary capsules $F_1\in\mathbb{R}^{\frac{C}{D_{i}} \times W\times H \times D_{i}}$, where $D_{i}$ is the dimensions of the primary capsules. In such design, there is $\frac{C}{D_{i}} \times W\times H$ number of capsules, each of capsule is in $\mathbb{R}^{D_{i}}$. Let $W_{ij} \in \mathbb{R}^{D_{i}\times [N \times R_{o}]}$ is a transformer matrix, each primary capsule $u_i$ is transformed to make a vote as $\hat{v}_{j|i} = W_{ij}u_i + B_j$, where $N$ is output classes and $D_{o}$ is the dimensions of output capsules. In CapsNet, a dynamic routing process at the $t^{th}$ iteration takes all votes into consideration to compute weight $c_{ij}$ for each vote $\tilde{u}_{j|i}$ as follows:

\begin{equation}
    \begin{split}
    &     s^t_j  = \sum_{i=1}^N{c_{ij}^t\hat{v}_{j|i}} \\ 
    &     v^t_j  = f^{squash}(s^t_j) \\ 
    &     c_{ij}^{t+1}  = \frac{exp(b_{ij} + \sum_{m=1}^t{v_j^m\hat{v}_{j|i})}}{\sum_{k=1}{exp(b_{ik} + \sum_{m=1}^tv_k^m\hat{v}_{k|i})}}\\
    & b_{ij}  = b_{ij} + \hat{v}_{j|i} v^t_j
    \end{split}
\label{eq:capsule}
\end{equation}
, $b_{ij}$ is log prior probability, $c_{ij}$ is coupling coefficient that models the degree with which $\hat{v}_{j|i}$ is able to predict $s_{j}$. It is initialized as $c_{ij} = \frac{exp(b_{ij})}{\sum_{k}{exp(b_{ik})}}$ .$f^{squash}$ is a squashing function that maps the length of a vector to [0, 1), i.e.,  
\begin{equation}
    v_j = f^{squash}(s) = \frac{|s|^2}{1+|s|^2}\frac{s}{|s|}
\end{equation}
The classification loss is defined as margin loss as follows:
\begin{equation}
    \mathcal{L}_{m} = \mathcal{I}_k max(0, m^+ - |v_k|)^2 + \lambda(1-\mathcal{I}_k)max(0, |v_k)|- m^-)^2
\end{equation}
As suggested by \cite{sabour2017dynamic}, $m^+, m^-, \lambda$ are set 0.9, 0.1 0.5. $\mathcal{I}_k =1$ if the object of the $k^{th}$ class is present in the input.

The reconstruction loss is computed as a regularization term in the loss function. The pseudo-code of the dynamic routing algorithm is presented in Algorithm \ref{al:routing}

\begin{algorithm}[t]
\caption{The pseudo-code of the dynamic routing algorithm.}
\label{al:routing}
\hrule
\begin{algorithmic}[1]
\algrenewcommand\algorithmicrequire{\textbf{Data: }}
\algrenewcommand\algorithmicensure{\textbf{Result: }}
\Require Capsule $v_i$ at layer (l).
\Ensure Capsule $v_j$ at layer (l+1).
\State Initial: for all capsule i in layer (l) and capsule j in layer (l + 1), $b_{ij} \gets 0$
\For{each iteration} 
    \For{all capsule i at layer (l)} $c_{ij} \gets \text{softmax}(b_{ij})$
    \EndFor
    \For{all capsule j at layer (l+1)} 
            \State $s_j  = \sum_{i=1}^N{c_{ij}^t\hat{v}_{j|i}}$ 
            \State $v_j  = f^{squash}(s_j)$
    \EndFor
    \For{all capsule i in layer (l) and capsule j in layer (l + 1)} 
        \State $b_{ij}  = b_{ij} + \hat{v}_{j|i} v_j$
    \EndFor
\EndFor
\end{algorithmic}
\end{algorithm}

Now, let consider the backpropagation through routing iterations. Assuming that there are $K$ iterations and $M$ capsules at output $v_1^k, v_2^K, ... v_M^K$, gradients through the routing procedure are:
\begin{equation}
    \frac{\partial\mathcal{L}}{\partial\hat{v}_{j|i}} = \frac{\partial\mathcal{L}}{\partial v_j^K}\frac{\partial v_j^K}{\partial s_j^K}c_{ij}^K + \sum_{m=1}^M{\frac{\partial\mathcal{L}}{\partial v_m^K}\frac{\partial v_m^K}{\partial s_m^K}\hat{v}_{m|i}\frac{\partial c_{im}^K}{\partial \hat{v}_{j|i}}}
\label{eq:backprop} 
\end{equation}
The second term in Eq.\ref{eq:backprop} is actually the main computational burden of the expensive routing.

We further investigate the robustness CapsNet when comparing it with CNNs as follows:
\begin{itemize}
    \item Translation invariant: While CNNs are able to identify if the object exists in a certain region, they are unable to identify the position of one object relative to another. Thus, CNNs can not model spatial relationships between objects/features. As shown in Fig. \ref{fig:CNN-translation}, CNNs can tell it is a dog image or face image but it can not tell the spatial relationship between the dog and the picture or position relation between facial components. 
    \begin{figure}[h]
    \centering
    \includegraphics[width=\textwidth]{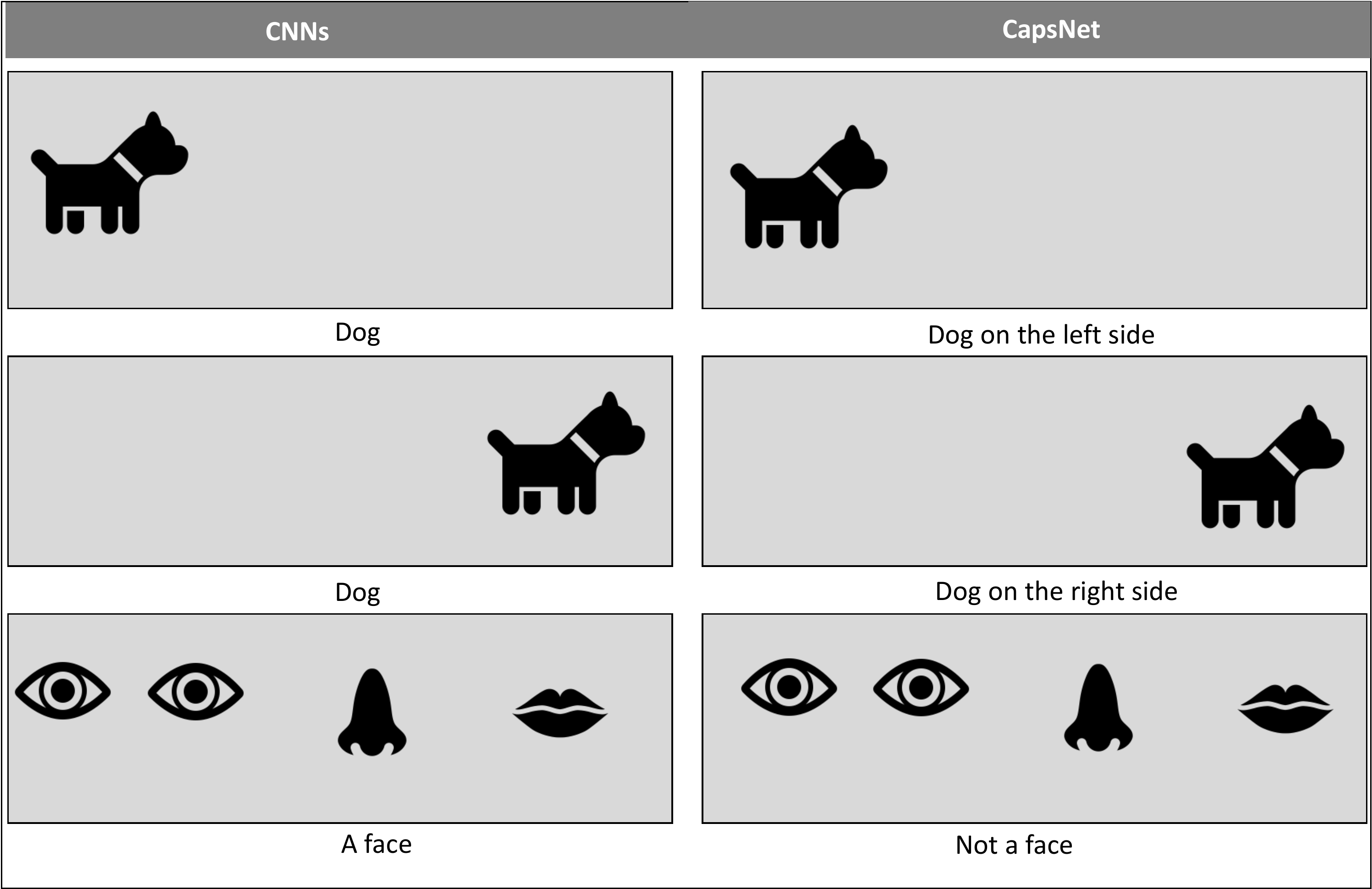}
    \caption{Translation invariant comparison between CNNs and CapsNet.}
    \label{fig:CNN-translation}
    \end{figure}
    \item Require less data to generalize: CNNs require to gether massive amounts of data that represents each object in various positions and poses. Then we train the CNNs on this huge dataset with a hope that the network is able to see enough examples of the object to generalize. Thus, to better generalize over variations of the same object, CNNs are trained on multiple copies of every sample, each being slightly different. Data augmentation is one of the most common techniques to make the CNNs model more robust. With CapsNet, it encodes invariant part-whole spatial relationships into learned weights. Thus, CapsNet is able to encode various positions and poses information of parts and the invariant part-whole relationships to generalize to unseen variations of the objects.
    \item Interpretability: There have been a large number of interpretation methods proposed to understand individual classifications of CNNs, model interpretability is still a significant challenge for CNNs. By taking part-whole relation into consideration, the higher capsule in CapsNet is interpretable and explainable. Thus, CapsNet is inherently more interpretable networks than traditional neural networks as capsules tend to encode specific semantic concepts. Especially, the disentangled representations captured by CapsNet often correspond to human understandable visual properties of input objects, e.g., rotations and translations. Let us take the example of reconstructing a MNIST digit from \cite{sabour2017dynamic}, different dimensions of the activity vector of a capsule controlled different features, including scale and thickness, localized part, stroke thickness, and width and translation as shown in Fig.\ref{fig:CNN-translation}.
    \begin{figure}[h]
    \centering
    \includegraphics[width=\textwidth]{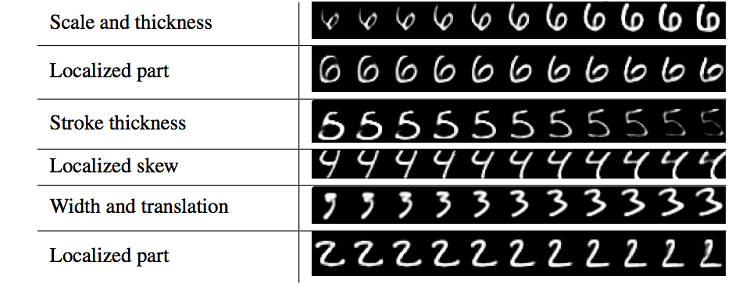}
    \caption{Experiments on MNIST. Different dimensions of the capsules are responsible for encoding different characteristics of the digits. Image is from \cite{sabour2017dynamic}.}
    \label{fig:CNN-translation}
    \end{figure}
    
\end{itemize}

\section{Capsule Network: Related Work}
There have been different mechanisms proposed to improve the performance of CapsNet. In general, we categorize them into two groups. The first one aims to propose various effective dynamic routing mechanisms to improve dynamic routing \cite{sabour2017dynamic}. Dynamic routing identifies the weights of predictions made by the lower-level capsules, called coupling coefficients by an iterative routing-by-agreement mechanism. EM Routing \cite{hinton2018matrix} updates coupling coefficients iteratively using Expectation-Maximization. By utilizing attention modules augmented by differentiable binary routers, \cite{chen2018generalized} proposes a straight-through attentive routing to reduce the high computational complexity of dynamic routing iterations in \cite{sabour2017dynamic}. To increase the computational efficiency, Mobiny, et al., \cite{mobiny2018fast} proposes a consistent dynamic routing mechanism that results in $3 \times$ speedup of CapsNet. Recently, \cite{ribeiro2020capsule} proposes a new capsule routing algorithm derived from Variational Bayes for fitting a mixture of transforming gaussians to show that it is possible transform capsule network into a Capsule-VAE. To reduce the parameters of CapsNet, \cite{hinton2018matrix, rajasegaran2019deepcaps} propose to use a matrix or a tensor to represent an entity instead of a vectors. The second category focuses on network architecture such as combinging both Convolutional layers and Capsule layers \cite{phaye2018multi}, unsupervised capsule autoencoder \cite{kosiorek2019stacked}, Aff-CapsNets \cite{gu2020improving}, Memory-augmented CapsNet\cite{mobiny2021memory} . While \cite{gu2020improving}  removes the dynamic routing by sharing the transformation matrix, \cite{gu2020interpretable} replaces the dynamic routing with a multi-head attention-based graph pooling approach to achieve better interpretability. Recently, Mobiny, et al.,  \cite{mobiny2020decaps} proposed DECAPS which utilize Inverted Dynamic Routing (IDR) mechanism to group lower-level capsules before sending them to higher-level capsules as well as employ a Peekaboo training procedure to encourage the network to focus on fine-grained information through a second-level attention scheme. DECAPS has outperformed experienced, well-trained thoracic radiologists \cite{mobiny2020radiologist}.

\section{CapsNets in Medical Image Segmentation}
This section introduces various recent CapsNets that have been proposed in medical image segmentation.
\subsection{2D-SegCaps}
\begin{figure}[h]
\centering
\includegraphics[width=\textwidth]{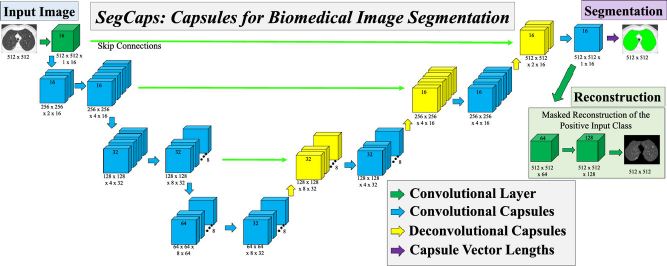}
\caption{Network architecture of 2D-SegCaps \cite{lalonde2018capsules, lalonde2021capsules} for biomedical image segmentation. The network is an UNet-based architecture with Capsule blocks are at both encoder and decoder paths.}
\label{fig:2DSegCaps}
\end{figure}

CapsNet has been mainly applied to image classification, image recognition; its performance is still limited compared to the state-of-the-art by CNNs-based approaches. 2D-SegCaps \cite{lalonde2018capsules, lalonde2021capsules} was the first CapsNet proposed for semantic image segmentation and illustrated as in Fig. \ref{fig:2DSegCaps}. As stated by \cite{lalonde2018capsules, lalonde2021capsules}, performing semantic image segmentation with a CapsNet is extremely challenging because of high computational complexity during the routing process which takes place between every parent and every possible child. The 2D-SegCaps is an UNet-based architecture with Capsule blocks are at both encoder and decoder paths. It contains four components corresponding to (i) visual feature extraction, which produces 16 feature maps of the same spatial dimensions, (ii) convolutional capsule at the encoder path, (iii) deconvolutional capsule at the decoder path, and (iv) reconstruction regularization at decoder path. Details of four components are as follows:
\begin{itemize}
    \item \textbf{Feature Extraction: }
    2D-SegCaps network takes a large 2D image $512 \times 512$ (e.g. a slice of a MRI scan) as its input. The image is passed through a 2D Conv layer which produces 16 feature maps of the same spatial dimensions, $512 \times 512 \times 16$. This becomes input of the following convolutional capsule layers.
    \item \textbf{Convolutional Capsule Encoder: }. The process of convolutional capsules and routing to any given layer $l$ in the network are given as follows:
    \begin{itemize}
    \item At layer $l$:
        There exists a set of capsule types \begin{equation}
            T^l = {t_1^l, t_2^l, ..., t_n^l, n\in \mathbb{N}}
        \end{equation}
        For every $t_i^l$, there exists an $h^l \times w^l$ grid of $d^l$-dimensional child capsules 
        \begin{equation}
        C = {c_{1, 1}, c_{1,2}, ... c_{1, w^l}, c_{2, 1}, ..., c_{h^l, w^l}}
        \end{equation}
        , where $h^l \times w^l$ is the spatial dimensions of the output of layer $l-1$.
    
    \item At layer $l+1$:
        There exists a set of capsule types
        \begin{equation}
            T^{l+1} = {t_1^{l+1}, t_2^{l+1}, ..., t_m^{l+1}, m\in \mathbb{N}}
        \end{equation}
        For every $t_i^{l+1}$, there exists an $h^{l+1} \times w^{l+1}$ grid of $d^{l+1}$-dimensional parent capsules 
        \begin{equation}
        P = {p_{1, 1}, p_{1,2}, ... p_{1, w^{l+1}}, p_{2, 1}, ..., c_{h^{l+1}, w^{l+1}}}
        \end{equation}
        , where $h^{l+1} \times w^{l+1}$ is the spatial dimensions of the output of layer $l$.
    \item For every parent capsule type $t_i^{l+1} \in T^{l+1}$, every parent capsule $p_{xy}$ receives a set of “prediction vectors”, each value is for each capsule type in $T^l$, i.e.,   $[\hat{v}_{{xy}|t_1^l}, \hat{v}_{{xy}|t_2^l}, ... \hat{v}_{{xy}|t_n^l}]$. This set of prediction vectors is defined as the matrix multiplication between a learned transformation matrix for the given parent capsule type, $M_{t_j^{l+1}}$, and the sub-grid of child capsules outputs, $V_{{xy}|t_i^l}$. The sub-grid of child capsules outputs  $V_{{xy}|t_i^l} \in \mathcal{R}^{h^k \times w^k \times d^l}$, where $h^k, w^k, d^l$ are the dimensions of the user-defined kernel, for all capsule types $t_i^l \in T^l$. Each parents capsules has dimension $M_{t_j^{l+1}} \in \mathcal{R}^{ h^k \times w^k \times d^l \times d^{l+1}}$. The
    $\hat{v}_{{xy}|t_i^l} \in \mathcal{R}^{d^{l+1}}$ is computed as:
    \begin{equation}
        \hat{v}_{{xy}|t_i^l} = M_{t_j^{l+1}} V_{{xy}|t_i^l}
    \end{equation}
    \end{itemize}
    To reduce total number of learned parameters, 2D-SegCaps shares transformation matrices across members of the grid, i.e., $M_{t_j^{l+1}}$ does not depend on the spatial location. This transformation matrix is shared across all spatial locations within a given capsule type. Such mechanism is similar to how convolutional kernels scan an input feature map and this is the main different between 2D-SegCaps and CapsNet. The parent capsule $p_{xy} \in P$ for parent capsule type $t_j^{l+1} \in T^{l+1}$ is then computed as follows:
    \begin{equation}
        p_{xy} \sum_{n}{c_{{t_i^l|xy}}}\hat{v}_{{xy}|t_i^l}
    \end{equation}
    where $c_{{t_i^l|xy}}$ is coupling coefficient defined in Eq.\ref{eq:capsule}, i.e., $c_{{t_i^l|xy}}  = \frac{exp\left(b_{t_i^l|xy}\right)} {\sum_{t_j^{l+1}}{exp\left(b_{t_i^{l}|t_j^{l+1}}\right)}}$ 
    The output capsule is then computed using a non-linear squashing function as defined in Eq.\ref{eq:capsule} as follows:
    \begin{equation}
        v_{xy} = f^{squash}(p_{xy})
    \end{equation}
    Lastly, the agreement is measured as the scalar product 
    \begin{equation}
        b_{xy} = v_{xy}\hat{v}_{xy|t_i^l}
    \end{equation}
    
    Unlike dynamic routing in CapsNet \cite{sabour2017dynamic}, 2D-SegCaps locally constrains the creation of the prediction vectors. Furthermore, 2D-SegCaps only routes the child capsules within the user-defined kernel $h^k \times w^k$ to the parent, rather than routing every single child capsule to every single parent.
    
    \item \textbf{Deconvolutional Capsule Decoder: }
    Deconvolutional capsules are as similar as the convolutional capsules; however, the prediction vectors are now formed using the transpose of the operation previously described. In the deconvolutional capsules, the set of prediction vectors are defined as the matrix multiplication between a learned transformation matrix, $M_{t_j^{l+1}}$ for a parent capsule type $t_j^{l+1} \in T^{l+1}$ and the sub-grid of child capsules outputs, $U_{xy|t_i^l}$ for each capsule type in $t_i^l \in T^{l}$. For each member of the grid, we can then form our prediction vectors again by the following equation.
    \begin{equation}
        \hat{u}_{xy|t_j^l} = M_{t_j^{l+1}}U_{xy|t_i^l}
    \end{equation}
    Similar to convolutional capsule encoder, $\hat{u}_{xy|t_j^l}$ is input to the dynamic routing algorithm to form our parent capsules and  $\hat{u}_{xy|t_j^l} \in \mathcal{R}^{d^{l+1}}$.
    \item \textbf{Reconstruction Regularization: }This component aims to model the distribution of the positive input class and treat all other pixels as background, the segmentation capsules which do not belong to the positive class is masked out. The reconstruction is performed via a three $1 \times 1$ Conv. layers. Then, it is computed by a mean-squared error (MSE) loss between only the positive input pixels and this reconstruction. The supervised loss for the reconstruction regularization is computed as follows:
    \begin{equation}
    \begin{split}
        R = I \times S|S \in \{0,1\} \\
        \mathcal{L}_{Reco} = \frac{\gamma}{H \times W} ||R - O||
    \end{split}
    \label{eq:res_loss}
    \end{equation}
    , where $I$ is the input image, $R$ is the reconstruction target, $S$ is the ground-truth segmentation mask, $O$ is the output of the reconstruction network. $\gamma$ is weighting coefficient for the reconstruction loss and set to 1 – 0.001. 
\end{itemize}

2D-SegCaps \cite{lalonde2018capsules, lalonde2021capsules} is trained with a supervised loss function. There are three loss functions included in the algorithm as follows:
\begin{itemize}
    \item \textbf{Margin Loss: } The margin loss is adopted from ~\cite{sabour2017dynamic} and it is defined between the predicted label $y$ and the ground truth label $y^*$ as follows:
    \begin{align}
    \mathcal{L}_{margin} =& y^* \times (\max(0, 0.9 - y))^2 + \\ & 0.5 \times (1 - y^*) \times (\max(0, y - 0.1))^2. \nonumber
    \end{align}
    Particularly, we compute the margin loss ($\mathcal{L}_{margin}$) on the capsule encoder output with downsampled ground truth segmentation. 
    
    \item \textbf{Weighted Cross Entropy Loss: }We compute the weighted cross-entropy loss ($\mathcal{L}_{CE}$) on the convolutional decoder. 
    
    \item \textbf{Reconstruction Regularization: } We also regularize the training with a network branch that aims at reconstructing the original input with masked mean-squared errors  ($\mathcal{L}_{Reco}$) as in Eq.\ref{eq:res_loss}. 
\end{itemize}

The total loss is the weighted sum of the three losses as follows:
\begin{equation}
    \mathcal{L} = \mathcal{L}_{margin} + \mathcal{L}_{CE} + \mathcal{L}_{Reco}
    \label{eq:loss}
\end{equation}

2D-SegCaps has obtained promising performance on LUNA16 dataset \cite{setio2017validation}; however, Survarachakan, et al. \cite{survarachakan2020capsule} has shown that 2D-SegCaps performance is significantly decreased on the MSD dataset \cite{simpson2019large} compared with Unet-based architectures. Furthermore, Survarachakan, et al. \cite{survarachakan2020capsule} extend 2D-SegCaps to Multi-SegCaps to support multiple class segmentation. Unlike 2D-SegCaps, the output capsule layer in Multi-SegCaps is modified to output $N 16D$ output capsules, where $N$ is the number of classes in the dataset, including background, and the predicted class is the one represented by the capsule with the longest euclidean length. Thus, Multi-SegCaps attempts to reconstruct the pixels belonging to all classes, except for the background class instead of a single target class as in 2D-SegCaps.

\begin{figure}[h]
\centering
\includegraphics[width=\textwidth]{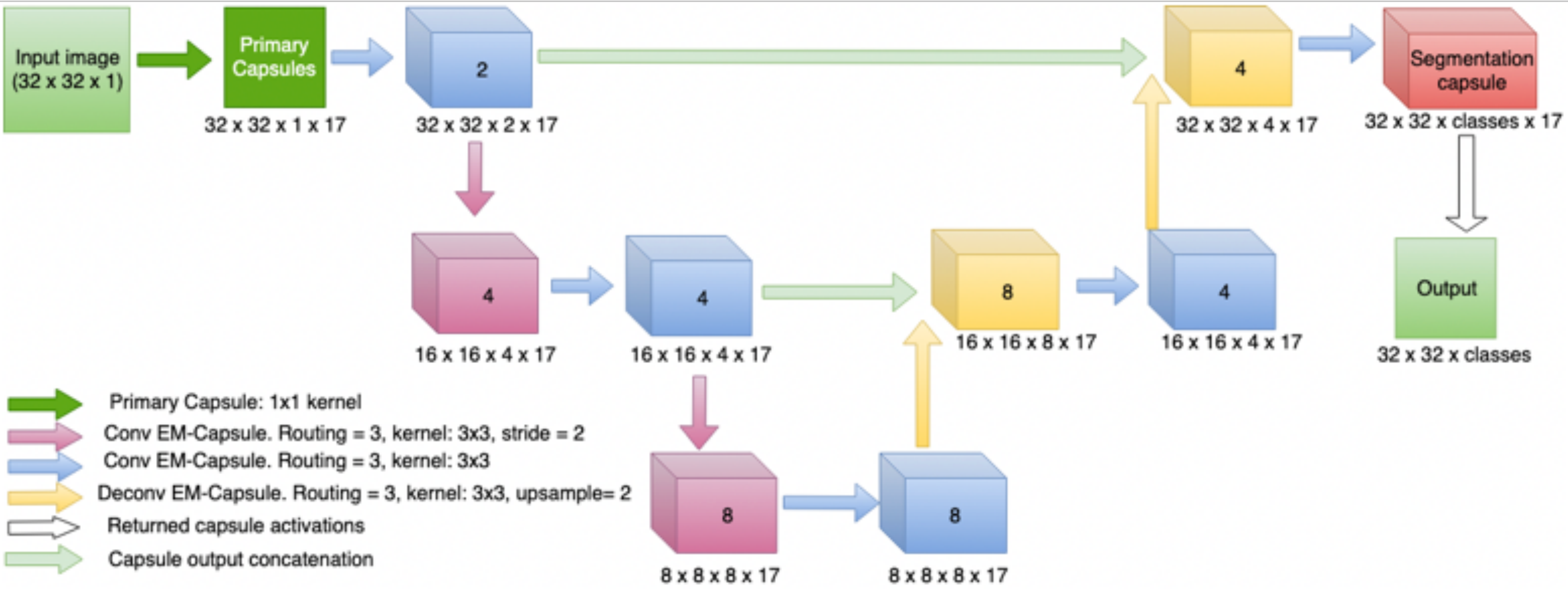}
\caption{EM-routing SegCaps architecture }
\label{fig:EM-routing-SegCaps}
\end{figure}

Like other CapsNet, 2D-SegCaps and Multi-SegCaps have the limitation of high computational complexity, which is caused by dynamic routing. To address such concern, Survarachakan, et al. \cite{survarachakan2020capsule} makes used on EM-routing \cite{hinton2018matrix} and proposes EM-routing SegCaps. The EM-routing SegCaps architecture uses matrix capsules with EM-routing and is shown in Fig.\ref{fig:EM-routing-SegCaps}. The main difference between EM-routing SegCaps and 2D-SegCaps is that convolutional capsule layers accept the poses and activations from capsules in the previous layer and output new poses and activations for the capsules in the next via Expectation-Maximization routing algorithm (EM-routing). In EM-routing SegCaps, all child capsules cast an initial vote of the output for every capsule in the next layer, using its own pose matrices before performance EM-routing. To cast this vote, a transformation matrix going into the parent capsule is trained and shared by all child capsules. In EM-routing SegCaps, predictions are first computed and then forwarded to the EM-routing algorithm, along with the activations from the previous layer. The EM-routing algorithm is run for three iterations before it returns the final pose and activations for all capsules in the current layer. By replacing dynamic routing in 2D-SegCaps by EM-routing, the performance of EM-routing SegCaps is not improved much compared with 2D-SegCaps. 

\subsection{3D-SegCaps}
2D-SegCaps has shown promising performance on 2D medical image segmentation, however, it ignores the temporal relationship when performing on 3D images, e.g. MRIs, CT scans. Nguyen, et al.\cite{nguyen20213d} extends 2D-SegCaps to 3D-SegCaps by incorporating temporal information into capsules. Similar to 2D-SegCaps, 3D-SegCaps contains four components corresponding to visual feature extraction, convolutional capsule layers, deconvolutional capsule layers, and reconstruction regularization. Details of four components are as follows:
\begin{itemize}
    \item \textbf{Feature Extraction: }
    3D-SegCaps network takes a volumetric data as its input. The volumetric data is passed through a 3D Conv layer which produces 16 feature maps of the same spatial dimensions, $64 \times 64 \times 64 \times 16$. This becomes input of the following convolutional capsule layers.
    \item \textbf{Convolutional Capsule Encoder: } The process of convolutional capsules and routing to any given layer $l$ in the network are given as follows:
    \begin{itemize}
    \item At layer $l$:
        There exists a set of capsule types \begin{equation}
            T^l = {t_1^l, t_2^l, ..., t_n^l, n\in \mathbb{N}}
        \end{equation}
        For every $t_i^l$, there exists an $h^l \times w^l \times r^l$ grid of $d^l$-dimensional child capsules 
        \begin{equation}
        \begin{split}
        C = \{c_{1, 1, 1}, c_{1,1,2}, ... c_{1,1, r^l}, c_{1, 2, 1}, ..., c_{1, w^l, 1}, c_{2, 1, 1}, ..., c_{h^l, 1, 1}, \\ 
        ... c_{1, w^l, r^l}, ..., c_{h^l, w^l, 1}, ... c_{h^l, 1, r^l}, ... c_{h^l, w^l, r^l}\}
        \end{split}
        \end{equation}
        where $h^l \times w^l \times r^l$ is the spatial dimensions of the output of layer $l-1$.
    
    \item At layer $l+1$:
        There exists a set of capsule types
        \begin{equation}
            T^{l+1} = {t_1^{l+1}, t_2^{l+1}, ..., t_m^{l+1}, m\in \mathbb{N}}
        \end{equation}
        For every $t_i^{l+1}$, there exists an $h^{l+1} \times w^{l+1} \times r^{l+1}$ grid of $d^{l+1}$-dimensional parent capsules 
        \begin{equation}
        \begin{split}
        P = \{p_{1, 1, 1}, p_{1,1,2}, ... p_{1,1, r^l}, p_{1, 2, 1}, ..., p_{1, w^l, 1}, p_{2, 1, 1}, ..., p_{h^l, 1, 1}, \\ 
        ... p_{1, w^l, r^l}, ..., p_{h^l, w^l, 1}, ... p_{h^l, 1, r^l}, ... p_{h^l, w^l, r^l}\}
        \end{split}
        \end{equation}
        where $h^{l+1} \times w^{l+1} \times r^{l+1}$ is the spatial dimensions of the output of layer $l$.
    \item For every parent capsule type $t_i^{l+1} \in T^{l+1}$, every parent capsule $p_{xyz}$ receives a set of “prediction vectors”, each value is for each capsule type in $T^l$, i.e.,   $[\hat{v}_{{xyz}|t_1^l}, \hat{v}_{{xyz}|t_2^l}, ... \hat{v}_{{xyz}|t_n^l}]$. This set of prediction vectors is defined as the matrix multiplication between a learned transformation matrix for the given parent capsule type, $M_{t_j^{l+1}}$, and the sub-grid of child capsules outputs, $V_{{xyz}|t_i^l}$. The sub-grid of child capsules outputs  $V_{{xyz}|t_i^l} \in \mathcal{R}^{h^k \times w^k \times r^k \times d^l}$, where $h^k, w^k, r^k, d^l$ are the dimensions of the user-defined kernel, for all capsule types $t_i^l \in T^l$. Each parents capsules has dimension $M_{t_j^{l+1}} \in \mathcal{R}^{ h^k \times w^k \times r^k \times d^l \times d^{l+1}}$. The
    $\hat{v}_{{xyz}|t_i^l} \in \mathcal{R}^{d^{l+1}}$ is computed as:
    \begin{equation}
        \hat{v}_{{xyz}|t_i^l} = M_{t_j^{l+1}} V_{{xyz}|t_i^l}
    \end{equation}
    \end{itemize}
    To reduce a total number of learned parameters, 3D-SegCaps shares transformation matrices across members of the grid i.e.,  $M_{t_j^{l+1}}$ does not depend on the spatial location. This transformation matrix is shared across all spatial locations within a given capsule type. Such a mechanism is similar to how convolutional kernels scan an input feature map and this is the main difference between 3D-SegCaps and CapsNet. The parent capsule $p_{xyz} \in P$ for parent capsule type $t_j^{l+1} \in T^{l+1}$ is then computed as follows:
    \begin{equation}
        p_{xyz} \sum_{n}{c_{{t_i^l|xyz}}}\hat{v}_{{xyz}|t_i^l}
    \end{equation}
    where $c_{{t_i^l|xyz}}$ is coupling coefficient defined in Eq.\ref{eq:capsule}, i.e.,  $c_{{t_i^l|xyz}}  = \frac{exp\left(b_{t_i^l|xyz}\right)} {\sum_{t_j^{l+1}}{exp\left(b_{t_i^{l}|t_j^{l+1}}\right)}}$ 
    Similar to 2D-SegCaps, the output capsule is then computed using a non-linear squashing function as defined in Eq.\ref{eq:capsule} as follows:
    \begin{equation}
        v_{xyz} = f^{squash}(p_{xyz})
    \end{equation}
    Finally, the agreement is measured as the scalar product 
    \begin{equation}
        b_{xyz} = v_{xyz}\hat{v}_{xyz|t_i^l}
    \end{equation}
    
    \item \textbf{Deconvolutional Capsule Decoder: }
    Deconvolutional capsules are similar to the one in 2D-SegCaps, in which the set of prediction vectors are defined again as the matrix multiplication between a learned transformation matrix, $M_{t_j^{l+1}}$ for a parent capsule type $t_j^{l+1} \in T^{l+1}$ and the sub-grid of child capsules outputs, $U_{xyz|t_i^l}$ for each capsule type in $t_i^l \in T^{l}$. For each member of the grid, we can then form our prediction vectors again by the following equation.
    \begin{equation}
        \hat{u}_{xyz|t_j^l} = M_{t_j^{l+1}}U_{xyz|t_i^l}
    \end{equation}
    where $\hat{u}_{xyz|t_j^l}$ is input to the dynamic routing algorithm to form our parent capsules and  $\hat{u}_{xyz|t_j^l} \in \mathcal{R}^{d^{l+1}}$.
    \item \textbf{Reconstruction Regularization}: This component is implemented in as the same manner as it is in 2D-SegCaps.
\end{itemize}

\subsection{3D-UCaps}
\begin{figure}[h]
\centering
\includegraphics[width=\textwidth]{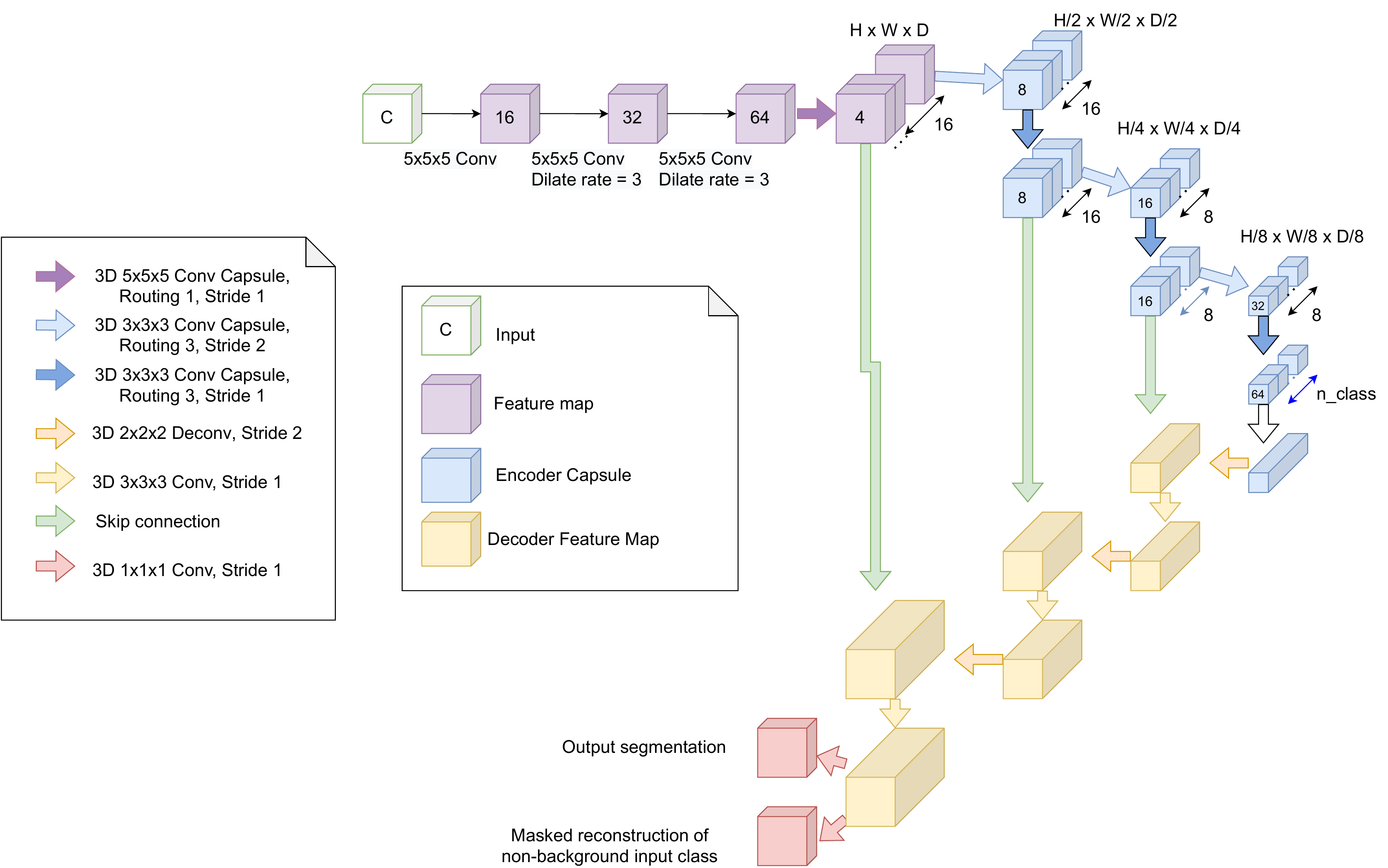}
\caption{3D-UCaps architecture with four components: visual feature extraction; convolutional capsule encoder, deconvolutional decoder, and reconstruction regularization. The number on the blocks indicates the number of channels in convolution layer and the dimension of capsules in capsule layers.}
\label{fig:3DUCaps}
\end{figure}

By taking temporal information into consideration, segmentation performance by 3D-SegCaps has been improved compared to 2D-SegCaps. However, the achievement by 3D-SegCaps is still lower than the SOTA performance by 3D-UNets. The observation by \cite{nguyen20213d} has shown that capsule design is capable of extracting richer representation comparing to traditional neural network design. Thus, convolutional capsule layer is utilized to encode visual representation in 3D-UCaps. Furthermore, under an encoder-decoder network architecture, the decoder path aims to produce a high detailed segmentation task, which has been high accurately performed by deconvolutional layers. Thus, deconvolutional layers are used at the decoder path in 3D-UCaps, which is the main difference between 3D-UCaps and 3D-SegCaps. This replacement does not only improve segmentation performance but also reduce computational cost caused by the dynamic routing. The entire network of 3D-UCaps is shown in Fig.\ref{fig:3DUCaps}. In CapsNet and CNNs. It follows Unet-like architecture \cite{cciccek20163d} and contains four main components as follows:

\begin{itemize}
    \item \textbf{Visual Feature Extractor:} A set of dilated convolutional layers is used to convert the input to high-dimensional features that can be further processed by capsules. It contains three convolution layers with the number of channels increased from 16 to 32 then 64, kernel size $5 \times 5 \times 5$ and dilate rate set to 1, 3, and 3, respectively. 
    The output of this part is a feature map of size $H \times W \times D \times 64$.
    \item \textbf{Convolutional Capsule Encoder:} This component is designed as similar mechanism as the one designed in 3D-SegCaps. The implement details of this component is as follows:
    The visual feature from the previous component can be cast (reshaped) into a grid of $H \times W \times D$ capsules, each represented as a single 64-dimensional vector. In the convolutional capsule encoder, it is suggested to be designed with more capsule types in low-level layers and less capsule types in high-level layers. 
    This is due to the fact that low-level layers represent simple object while high-level layers represent complex object and the clustering nature of routing algorithm \cite{hinton2018matrix}. 
    The number of capsule types in the encoder path of our network are set to $(16, 16, 16, 8, 8, 8)$, respectively. 
    This is in contrast to the design in 2D-SegCaps and 3D-SegCaps where the numbers of capsules are increasing $(1, 2, 4, 4, 8, 8)$ along the encoder path.
    The number of capsule types in the last convolutional capsule layer is equal to the number of categories in the segmentation, which can be further supervised by a margin loss~\cite{sabour2017dynamic}. 
    The output from a convolution capsule layer has the shape $H \times W \times D \times C \times A$, where $C$ is the number of capsule types and $A$ is the dimension of each capsule. 
    \item \textbf{Deconvolutional Decoder:} The decoder of 3D Unet \cite{cciccek20163d} is used in expanding path. This contains deconvolution, skip connection, convolution and BatchNorm layers \cite{ioffe2015batch} to generate the segmentation from features learned by capsule layers.The features is reshaped to $H \times W \times D \times (C \star A)$ before passing them to the next convolution layer or concatenating with skip connections. 
    \item \textbf{Reconstruction Regularization :} This component is implemented in as the same manner as it is in 3D-SegCaps.
\end{itemize}

\subsection{SS-3DCapsNet}
Despite the recent success of CapsNet-based approaches in medical image segmentation, there remains a wide range of challenges: 
(1) Most methods are based on supervised learning, which is prone to many data problems like small-scale data, low-quality annotation, small objects, ambiguous boundaries, to name a few. 
These problems are not straightforward to overcome: labeling medical data is laborious and expensive, requiring an expert's domain knowledge.
(2) Capsule networks for medical segmentation do not outperform CNNs yet, even though the performance gap gets significantly closer~\cite{nguyen20213d}.

To address the aforementioned limitations, Tran et al., \cite{tran2022ss} improve 3D-UCaps and propose SS-3DCapsNet, a self-supervised capsule network. Self-supervised learning (SSL) is a technique for learning feature representation in a network without requiring a labeled dataset. 
A common workflow to apply SSL is to train the network in an unsupervised manner by learning with a pretext task in the pre-training stage, and then fine-tuning the pre-trained network on a target downstream task. In the case of MIS, the suitable pretext tasks can be considered in four categories: context-based, generation-based, free semantic label-based, and cross-modal-based. In SS-3DCapsNet \cite{tran2022ss}, the pretext task is based on image reconstruction. 

\begin{figure}[h]
\centering
\includegraphics[width=\textwidth]{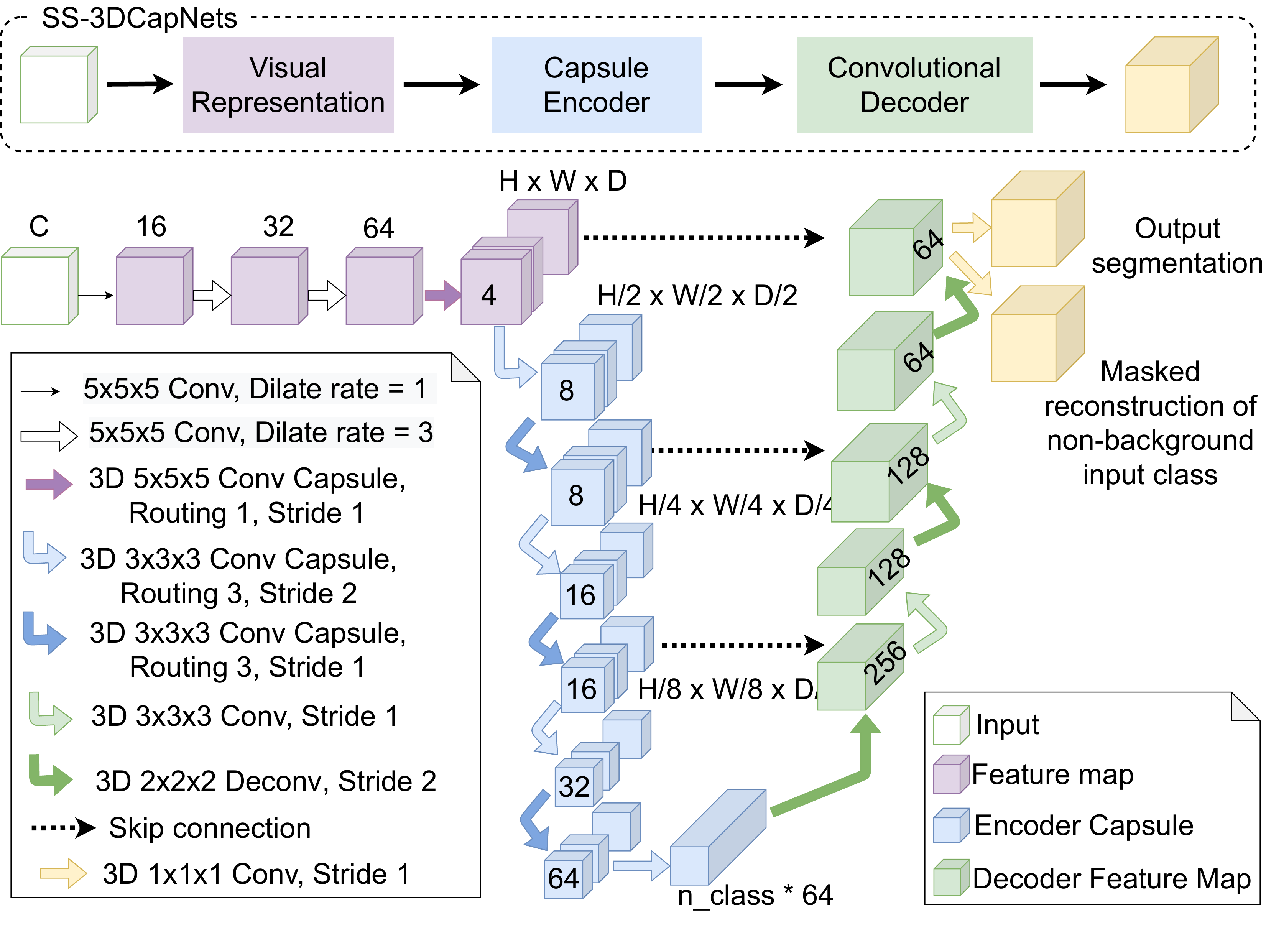}
\caption{SS-3DCapsNet architecture with three components: visual representation; convolutional capsule encoder, and deconvolutional decoder. Number on the blocks indicates number of channels in convolution layer and dimension of capsules in capsule layers.}
\label{fig:3DUCaps}
\end{figure}

The pretext task and downstream task in SS-3DCapsNet are detailed as follows:
\begin{itemize}
    \item \textbf{Pretext Task: } In computer vision, it is common to use pseudo-labels defined by different image transformations, e.g, rotation, random crop, adding noise, blurring, scaling, flipping, jigsaw puzzle, etc. to supervise the pretext task.  
    While such transformations work well for classification as a downstream task, they can not be applied directly into image segmentation. In SS-3DCapsNet \cite{tran2022ss}, image reconstruction from various transformations, i.e.,  noisy, blurring, zero-channels (R,G,B), swapping as shown in Fig. \ref{fig:transformation} is utilized to perform pretext task.

    \begin{figure}[t]
    \centering
    \includegraphics[width=\linewidth]{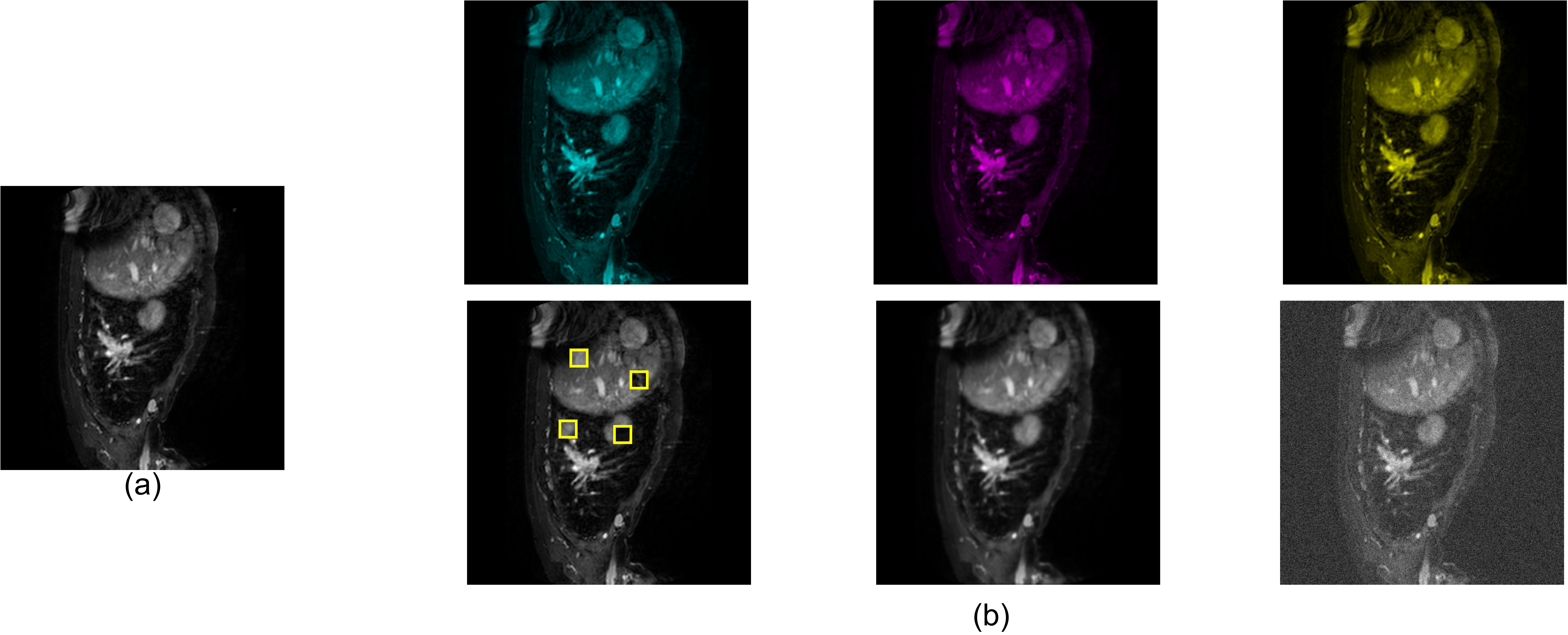}
    \caption{Examples of six transformations for self-supervised learning. (a): original image. (b) from left to right, top to bottom: zeros-green-channel, zeros-red-channel, zeros-blue-channel, swapping (4 swapped patches are shown in yellow boxes), blurring, noisy.}
    
    \label{fig:transformation}
\end{figure}

\begin{figure}[t]
    \centering
    \includegraphics[width=\textwidth]{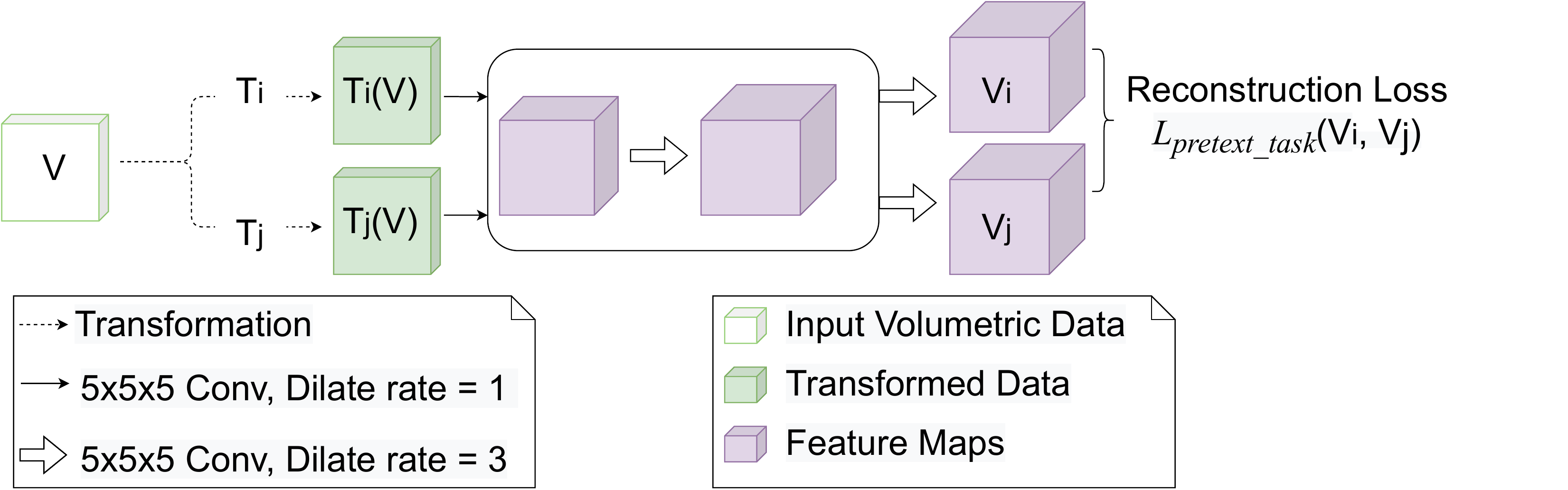}
    \caption{Our pretext task with reconstruction loss. \vspace{-5mm}}
    \label{fig:SSL}\
\end{figure}

Let $\mathcal{F}$ is the visual representation network. The transformation is defined as $\{T_i\}_{i=1}^{i=N}$, where $T_0$ is an identity transformation and $N$ is set as 6 corresponding to six transformations (Fig. \ref{fig:transformation}). 
Let $V$ denote as the original input volumetric data. The pretext task is performed by applying two random transformations $T_i, T_j (i, j \in [0, 6])$ into $V$. The transformed data is then $T_i(V)$ and $T_j(V)$, respectively. The visual feature of transformed data after applying the network $\mathcal{F}$ is $V_j$ and $V_j$, where $V_i = \mathcal{F}(T_i(V))$ and $V_j = \mathcal{F}(T_j(V))$. 
The network $\mathcal{F}$ is trained with a reconstruction loss defined by:
\begin{equation}
    \mathcal{L}_{pretext}(V_i, V_j) = ||V_i - V_j||_{2}.
\end{equation}
The pretext task procedure is illustrated in Fig. \ref{fig:SSL}.

\item \textbf{Downstream Task: } After pre-training, SS-3DCapsNet network is trained with annotated data on the medical segmentation task. The total loss function to train this downstream task is a sum of three losses, i.e.,  margin loss, weighted cross-entropy loss and reconstruction regularization loss as defined in Eq. \ref{eq:loss}

\end{itemize}

\subsection{Comparison}
In this section, the comparison will be conducted regarding both network architecture and performance accuracy.

The network architecture comparison between various CapsNet-based approaches for medical image segmentation  is shown in Table \ref{tab:compare}.

\begin{table*}[h]
\caption{Network architecture comparison between various CapsNet-based image segmentation }
\label{tab:compare}
\begin{tabular}{lllll}
\toprule
        & Input               & \shortstack{Initialization}      & Encoder   & Decoder    \\ \midrule
2D-SegCaps~\cite{lalonde2018capsules,lalonde2021capsules}& 2D still image             & Random                    & Capsule  & Capsule \\
3D-SegCaps~\cite{nguyen20213d}& 3D volumetric       & Random                    & Capsule  & Capsule \\
3D-UCaps~\cite{nguyen20213d} & 3D volumetric & Random & Capsule  & Deconvolution \\
 SS-3DCapsNet\cite{tran2022ss}  & 3D volumetric & SSL & Capsule & Deconvolution \\ \bottomrule
\end{tabular}
\end{table*}

\begin{figure}[h]
    \centering
    \includegraphics[width=0.5\textwidth]{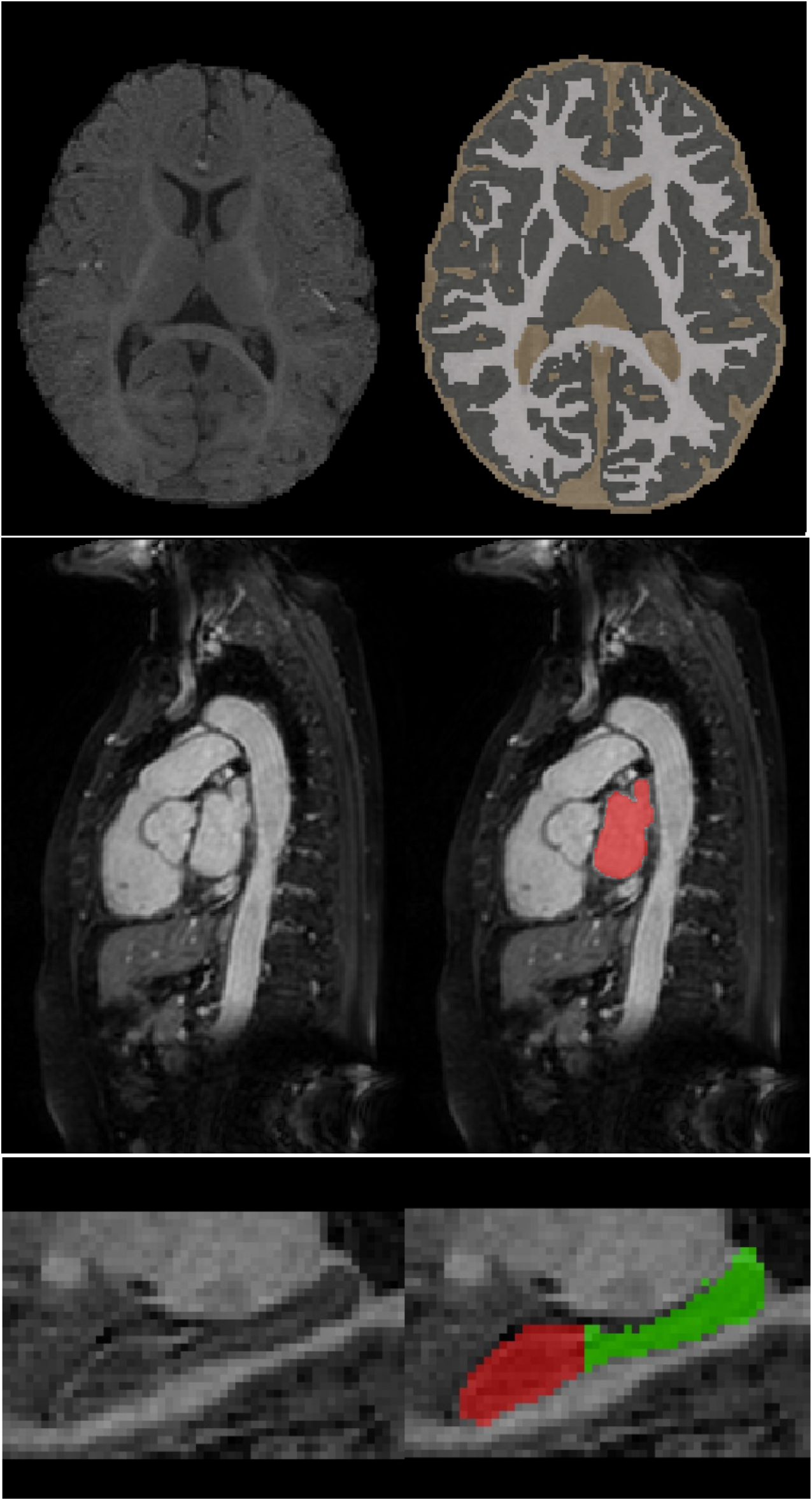}
    \caption{Visualization of samples from iSeg (first row), Cardiac (second row), and Hippocampus (third row).}
    \label{fig:dataset}
\end{figure}

To compare the performance of various CapNets-based approaches, small-size datasets such as iSeg \cite{wang2019benchmark}, Cardiac, and Hippocampus~\cite{simpson2019large} are selected to conduct experimental results. Samples from three datasets are visualized in Fig.\ref{fig:dataset}.

\begin{itemize}
    \item \textbf{iSeg dataset:}\cite{wang2019benchmark} is an infant brain dataset consisting of 10 subjects with ground-truth labels for training and 13 subjects without ground-truth labels for testing.  Subjects were selected from Baby Connectome Project \cite{BCP} and have average age is in [5.5 - 6.5] months at the time of scanning. Each subject includes T1-weighted and T2-weighted images with size of $144 \times 92 \times 256$ and image resolution of $1 \times 1 \times 1 mm$. The difficulty of this dataset lies in the low contrast between tissues in the infant brain MRI that can reduce the accuracy of the automatic segmentation algorithms.
    \item \textbf{Cardiac: }\cite{simpson2019large}
    is a mono-modal MRI dataset containing 20 training images and 10 testing images covering the entire heart acquired during a single cardiac phase. This dataset was provided by King's College London and obtained with voxel resolution $1.2 \times 1.25 \times 2.75 mm^3$.
    \item \textbf{Hipposcampus: }
    \cite{simpson2019large} is a larger-scale mono-modal MRI dataset taken from the Psychiatric Genotype/Phenotype Project data repository at Vanderbilt University Medical Center (Nashville, TN, USA). It consists of 260 training and 130 testing samples acquired with a 3D T1-weighted MPRAGE sequence (TI/TR/TE,
    860/8.0/3.7 ms; 170 sagittal slices; voxel size, $1 \times 1 \times 1 mm^3$ ). The task of this dataset is segmenting two neighbouring small structures (posterior and anterior hippocampus) with high precision.
\end{itemize}

\begin{table}[!h]
\small
\centering
\caption{Performance comparison on iSeg-2017.}
\label{table:iseg}
\begin{tabular}{@{}ll  llll@{}}
\toprule
\multicolumn{1}{c}{\multirow{2}{*}{Method}} & \multicolumn{1}{c}{\multirow{2}{*}{Depth}} & \multicolumn{4}{c}{Dice Score}  \\    &             & \multicolumn{1}{c}{WM}             & \multicolumn{1}{c}{GM}             & \multicolumn{1}{c}{CSF} & Average        \\ \midrule                           \\  
2D-SegCaps \cite{lalonde2018capsules}                                  & 16      & 82.80                              & 84.19                              & 90.19                   & 85.73          \\
3D-SegCaps \cite{nguyen20213d}     & 16   & 86.49                              & 88.53                              & 93.62                   & 89.55  \\
3D-UCaps \cite{nguyen20213d}  & 17 & 90.21 & 91.12 &  {94.93} & 92.08\\

Our SS-3DCapsNet \cite{tran2022ss}                      & 17     &  {90.78}                             &  {91.48}                             & 94.92                   & {92.39}         \\ \bottomrule
\end{tabular}
\end{table}

\begin{table}[h]
\centering
\caption{Comparison on Cardiac with 4-fold cross validation.}
\label{table:cardiac}
\centering
\begin{tabular}{ll}
\toprule
 SegCaps (2D)~\cite{lalonde2018capsules}   &  66.96 \\ 
 Multi-SegCaps (2D)~\cite{survarachakan2020capsule}  & 66.96 \\
3D-UCaps \cite{nguyen20213d} & 89.69 \\ 
 SS-3DCapsNet \cite{tran2022ss}  &        89.77         \\  
\bottomrule
\end{tabular}
\end{table}

\begin{table}[h]
\centering
\caption{Comparison on Hippocampus with 4-fold.}
\label{table:hippocampus}
\vspace*{0.2cm}
\resizebox{1.0\linewidth}{!}{
\begin{tabular}{@{}llll|lll@{}}
\toprule
\multicolumn{1}{c}{Method} & Anterior &           &        & Posterior &           &        \\ \cmidrule(l){2-7} 
\multicolumn{1}{c}{}       & Recall   & Precision & Dice   & Recall    & Precision & Dice   \\ \midrule
Multi-SegCaps (2D)~\cite{survarachakan2020capsule}               & 80.76   & 65.65    & 72.42 & 84.46    & 60.49    & 70.49 \\
EM-SegCaps (2D)~\cite{survarachakan2020capsule}                  & 17.51   & 20.01    & 18.67 & 19.00    & 34.55    & 24.52 \\
3D-UCaps \cite{nguyen20213d}                       & 81.70  & 80.19  &80.99 &  80.2 & 79.25 &79.48  \\ 

SS-3DCapsNet \cite{tran2022ss}                       & {81.84}  &  {81.49 }  &  {81.59} &   {80.71 } &  {80.21 }   & {79.97} \\ \bottomrule
\end{tabular}
}
\end{table}

\begin{figure}[h!]
    \centering
    \includegraphics[width=\textwidth]{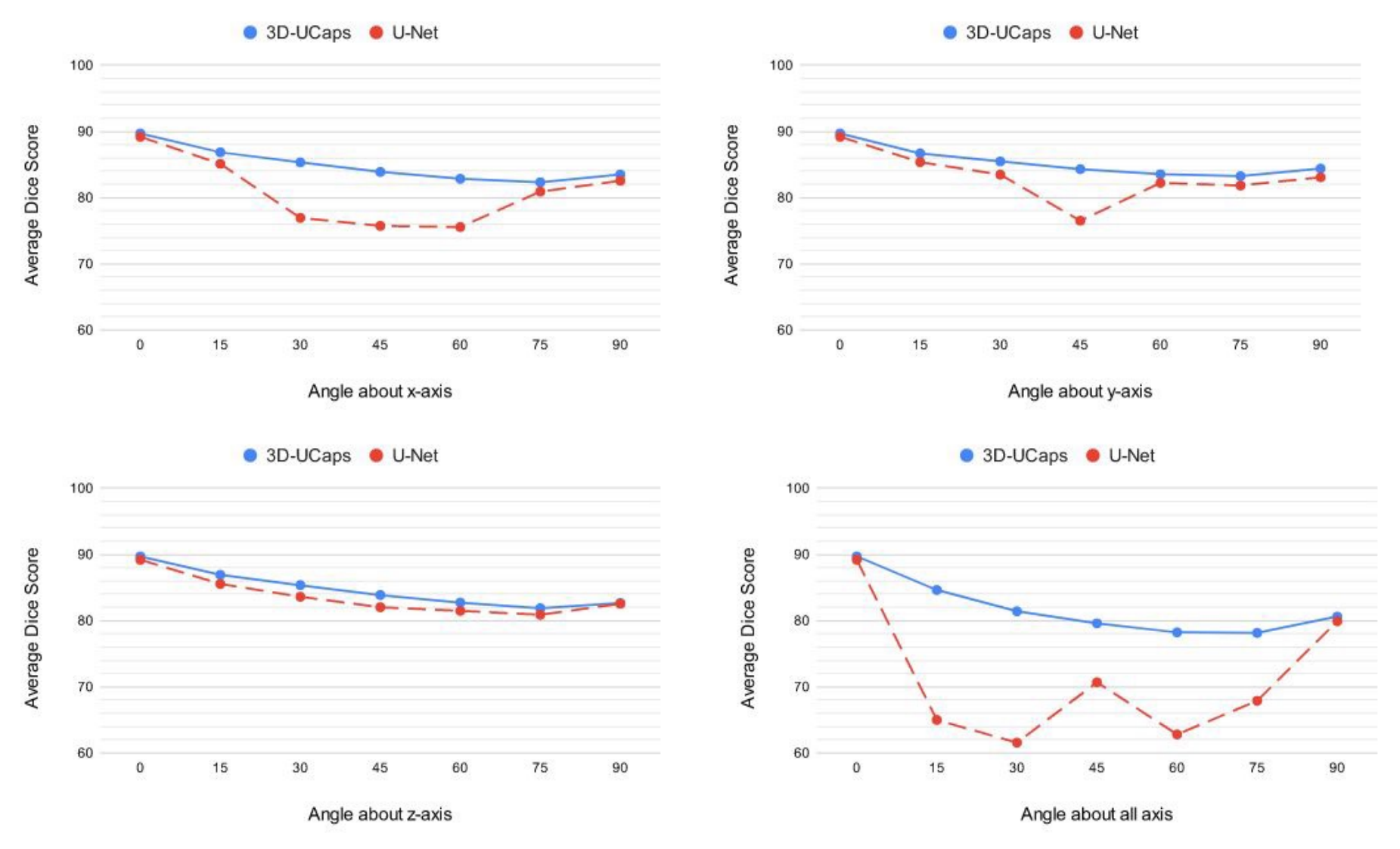}
    \caption{Performance comparison  on iSeg of 3D-UCaps and 3D-UNet with rotation equivariance on x, y, z, and all axis.}
    \label{fig:xyzrotation}
\end{figure}

\begin{figure}[h]
    \centering 
    \includegraphics[width=\textwidth]{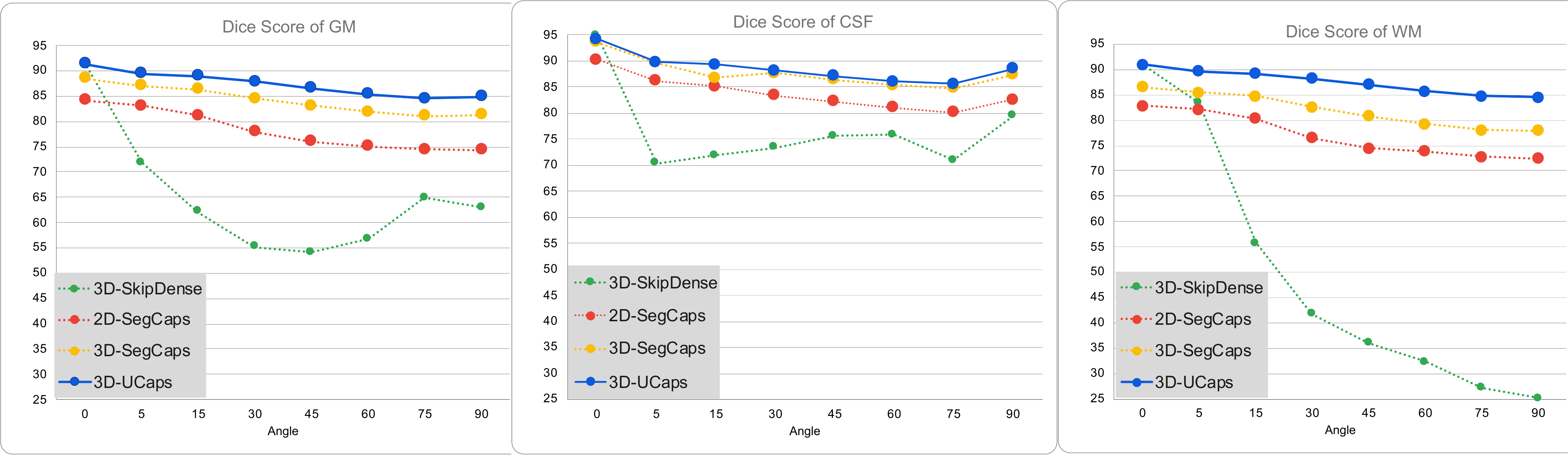}
    \caption{Performance comparison on iSeg with various various network on a particular axis rotation equivariance.}
    \label{fig:zrotation}
\end{figure}

\begin{figure}[h!]
    \centering
    \includegraphics[width=0.8\textwidth]{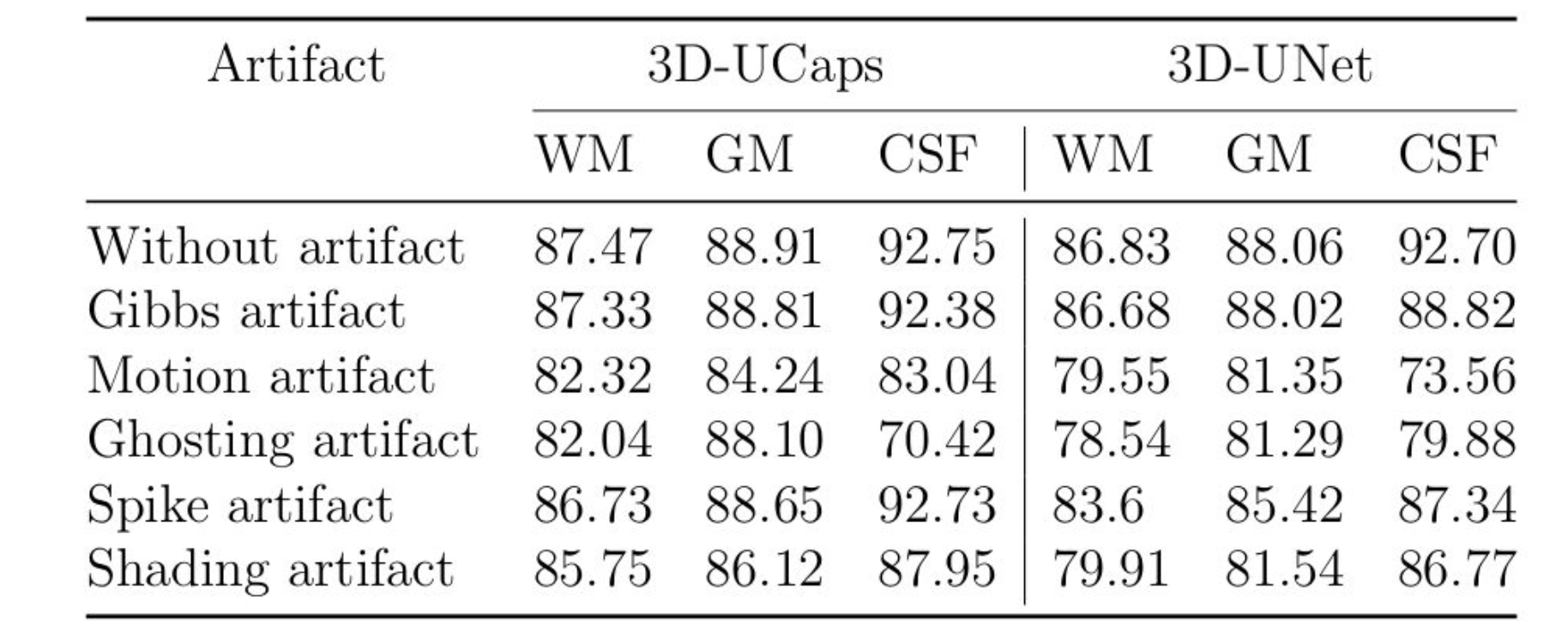}
    \caption{Performance comparison between 3D-UCaps and 3D-UNet on iSeg with various artifact.}
    \label{fig:artifact}
\end{figure}
For iSeg, the experimental results is followed by \cite{bui2019skip} in which of 9 subjects are used to train and subject \#9 is used to test. On Cardiac, and Hippocampus~\cite{simpson2019large}, the experiments are conducted by 4-fold cross-validation.  

The comparison is conducted by Pytorch. Patch size are selected as follows: $64 \times 64 \times 64$ for iSeg and Hippocampus, $128 \times 128 \times 128$ for Cardiac. All the networks were trained without any data augmentation. Adam optimizer with an initial learning rate of 0.0001 is chosen. The learning rate is decayed by 0.05 if the Dice score on the validation set does not increase for 50,000 iterations. Early stopping is set at 250,000 iterations as in \cite{lalonde2018capsules}. 

The performance comparison on various CapsNet-based medical image segmentation approaches is shown in Table.\ref{table:iseg}, Table.\ref{table:cardiac}, Table.\ref{table:hippocampus} corresponding to iSeg, Cardiac and Hippocampus datasets.

\section{Discussion}

Although CNNs have achieved outstanding performance on various tasks including medical image segmentation, they suffer from the loss of part-whole relationships and geometric information. CapsNet was proposed to address such limitations. 3D-UCaps \cite{nguyen20213d} conducted an analysis with two experiments on small-size datasets iSeg with rotation equivariance and invariance properties to various artifact as follows:
\begin{itemize}
    \item \textbf{Rotation Equivariance}: In the first experiment, the testing subject is rotated from 0 to 90 degrees (15, 30, 45, 60, 75, 90) on  x-axis, y-axis, z-axis, all-axis. The performance comparison on rotation equivariance between 3D-SegCaps and 3D-UNet is shown in Fig.\ref{fig:xyzrotation}. 
    Furthermore, the performance comparison between various networks, i.e., 3D-UCaps, 3D-SegCaps, 2D-SegCapsand, and 3D-UNet on a particular axis , i.e.,  z-axis is shown in Fig.\ref{fig:zrotation}.

    \item \textbf{Various Artifact}: In the second experiment, MonAI \cite{MonAI} and TorchIO \cite{perez2021torchio} are utilized to create artifacts. The performance comparison on i-Seg between 3D-UCaps and 3D-UNet is shown in Fig. \ref{fig:artifact}

CapsNets with their capability of modelling the part-whole relationships have obtained remarkable results various tasks including medical image segmentation. The aforementioned discussion has proved that CapsNets significantly outperform CNNs for small-size datasets, which is a common case in medical image segmentation applications due to the lack of annotated data. The experimental results also show that CapsNets obtain higher robustness to affine transformations than CNNs, however, their performances are still limited on unseen transformed inputs and their computational complexity is still high. Exploring hybrid architecture between CapsNet-based and traditional neural network is therefore a promising approach to medical image analysis while keeping model complexity and computation cost plausible.

\end{itemize}

\section*{Acknowledgment}
This material is based upon work supported by the National Science Foundation under Award No. OIA-1946391.

\printbibliography

\end{document}